%% file: neurips_2024.tex
\documentclass{article}

\usepackage[final]{neurips_2024}

\usepackage[utf8]{inputenc} 
\usepackage[T1]{fontenc}    
\usepackage{hyperref}       
\usepackage{url}            
\usepackage{booktabs}       
\usepackage{amsfonts}       
\usepackage{nicefrac}       
\usepackage{caption}
\usepackage{microtype}      
\usepackage{xcolor}         
\usepackage{graphicx, wrapfig}
\usepackage{amsmath}
\usepackage{enumitem}
\usepackage{multirow}
\usepackage{cleveref}
\usepackage{algorithm}
\usepackage{algorithmic}

\title{Optimistic Verifiable Training by Controlling Hardware Nondeterminism}

\author{%
 Megha Srivastava\thanks{Correspondence to \url{megha@cs.stanford.edu}. While our work focuses on training verification, our method to control hardware nondeterminism may be useful for other applications, such as repeated inference for text generation, as well as general concerns of machine learning reproducibility. } \\
  Department of Computer Science\\
  Stanford University\\
  \texttt{megha@cs.stanford.edu} \\
  \And
  Simran Arora \\
  Department of Computer Science \\
  Stanford University \\
  \texttt{simarora@stanford.edu} \\
  \And
  Dan Boneh \\
  Department of Computer Science \\
  Stanford University \\
  \texttt{dabo@cs.stanford.edu} \\
}

\begin{document}

\maketitle

\begin{abstract}
The increasing compute demands of AI systems have led to the emergence of services that train models on behalf of clients lacking necessary resources. However, ensuring correctness of training and guarding against potential training-time attacks, such as data poisoning and backdoors, poses challenges. Existing works on verifiable training largely fall into two classes: proof-based systems, which are difficult to scale, and ``optimistic'' methods that consider a third-party auditor who can replicate the training process and contest the trainer. A key challenge with the latter is that nondeterminism between GPU types during training prevents exact replication of the training process, resulting in schemes that are non-robust. We propose a method that combines training in a higher precision than the target, rounding after intermediate computations, and sharing rounding decisions based on an adaptive thresholding procedure, to successfully control for nondeterminism.  Across three different NVIDIA GPUs (A40, Titan XP, RTX 2080 Ti), we achieve exact training replication at FP32 precision for both full-training and fine-tuning of ResNet-50 (23M) and GPT-2 (117M) models. Our  verifiable training scheme significantly decreases the storage and time costs compared to proof-based systems, and is publicly released at \url{https://github.com/meghabyte/verifiable-training}.
\end{abstract}

\section{Introduction}
\input{sections/introduction} 

\section{Related Works} 
\input{sections/related_work}

\section{Set-Up: The Verification Game}
\input{sections/setup}

\input{sections/challenge}

\section{Method Overview}
\input{sections/method}

\section{Empirical Results}
\input{sections/experiments}

\section{Security Analysis}
\input{sections/security}

\section{Limitations and Future Work}\label{sec:discussion}

Our verifiable training scheme successfully controls for hardware nondeterminism. It expands the pool of potential auditors of a model training service, allowing us to envision a world where a $\mathrm{client}$ can even use two competing service providers it trusts to audit each other. Relative to proof-based systems, a limitation is the need for all parties to trust the third-party $\mathrm{auditor}$. If the $\mathrm{trainer}$ provides finetuning services on top of closed-source models (e.g., OpenAI), then our scheme will only work for the third-party auditors that the $\mathrm{trainer}$ is willing to share model weights with.  Other limitations included the added latency of training in higher precision and the storage cost. While we have shown that our method requires significantly less storage than alternatives, the vast majority of stored rounding decisions are not used by the $\mathrm{auditor}$ and are therefore unnecessary (Section \ref{sec:experiments}). Therefore, an exciting direction for future work is to mitigate this gap by better predicting when GPU divergence between computations occurs. Recent work has similarly argued for a stronger profile of noise during training in the context of verification \citep{fang2023proofoflearning}. Finally, another direction for future work includes adapting our method for distributed training \citep{li2020vldb}. 

\section{Acknowledgements}
We thank Bill Dally, Duncan Riach, Gabriel Poesia, and Chris Ré for helpful discussion and feedback. Megha Srivastava was supported by an IBM PhD Fellowship and the NSF Graduate Research Fellowship Program under Grant No. DGE-1656518.  In addition, this work was funded by NSF, DARPA, the Simons Foundation, UBRI, and NTT Research.  Opinions, findings, and conclusions or recommendations expressed in this material are those of the authors and do not
necessarily reflect the views of DARPA.

\bibliographystyle{unsrtnat}
{
\small
\bibliography{custom}
}


\appendix
\input{sections/appendix}

\clearpage
\newpage
\newpage
\newpage
\newpage
\section*{NeurIPS Paper Checklist}

\begin{enumerate}

\item {\bf Claims}
    \item[] Question: Do the main claims made in the abstract and introduction accurately reflect the paper's contributions and scope?
    \item[] Answer: \answerYes{}
    \item[] Justification: See abstract and introduction.

\item {\bf Limitations}
    \item[] Question: Does the paper discuss the limitations of the work performed by the authors?
    \item[] Answer:\answerYes{}
    \item[] Justification:  See Limitations.

\item {\bf Theory Assumptions and Proofs}
    \item[] Question: For each theoretical result, does the paper provide the full set of assumptions and a complete (and correct) proof?
    \item[] Answer: \answerNA{}
    \item[] Justification: \answerNA{}

    \item {\bf Experimental Result Reproducibility}
    \item[] Question: Does the paper fully disclose all the information needed to reproduce the main experimental results of the paper to the extent that it affects the main claims and/or conclusions of the paper (regardless of whether the code and data are provided or not)?
    \item[] Answer:\answerYes{}
    \item[] Justification: See Experiments and Appendix.

\item {\bf Open access to data and code}
    \item[] Question: Does the paper provide open access to the data and code, with sufficient instructions to faithfully reproduce the main experimental results, as described in supplemental material?
    \item[] Answer: \answerYes{}
    \item[] Justification: Code will be released in public version.

\item {\bf Experimental Setting/Details}
    \item[] Question: Does the paper specify all the training and test details (e.g., data splits, hyperparameters, how they were chosen, type of optimizer, etc.) necessary to understand the results?
    \item[] Answer: \answerYes{}
    \item[] Justification: See Experiments and Appendix.

\item {\bf Experiment Statistical Significance}
    \item[] Question: Does the paper report error bars suitably and correctly defined or other appropriate information about the statistical significance of the experiments?
    \item[] Answer: \answerNA{}
    \item[] Justification: No experiment results have variation necessitating statistical significance.

\item {\bf Experiments Compute Resources}
    \item[] Question: For each experiment, does the paper provide sufficient information on the computer resources (type of compute workers, memory, time of execution) needed to reproduce the experiments?
    \item[] Answer: \answerYes{}
    \item[] Justification: See Appendix for GPUs.
    
\item {\bf Code Of Ethics}
    \item[] Question: Does the research conducted in the paper conform, in every respect, with the NeurIPS Code of Ethics \url{https://neurips.cc/public/EthicsGuidelines}?
    \item[] Answer: \answerYes{}
    \item[] Justification: Yes

\item {\bf Broader Impacts}
    \item[] Question: Does the paper discuss both potential positive societal impacts and negative societal impacts of the work performed?
    \item[] Answer: \answerYes{}
    \item[] Justification: See Security Analysis.
    
\item {\bf Safeguards}
    \item[] Question: Does the paper describe safeguards that have been put in place for responsible release of data or models that have a high risk for misuse (e.g., pretrained language models, image generators, or scraped datasets)?
    \item[] Answer: \answerNA{}
    \item[] Justification: \answerNA{}

\item {\bf Licenses for existing assets}
    \item[] Question: Are the creators or original owners of assets (e.g., code, data, models), used in the paper, properly credited and are the license and terms of use explicitly mentioned and properly respected?
    \item[] Answer: \answerNA{}
    \item[] Justification: \answerNA{}

\item {\bf New Assets}
    \item[] Question: Are new assets introduced in the paper well documented and is the documentation provided alongside the assets?
    \item[] Answer: \answerNA{}
    \item[] Justification: \answerNA{}

\item {\bf Crowdsourcing and Research with Human Subjects}
    \item[] Question: For crowdsourcing experiments and research with human subjects, does the paper include the full text of instructions given to participants and screenshots, if applicable, as well as details about compensation (if any)? 
    \item[] Answer: \answerNA{}
    \item[] Justification: \answerNA{}

\item {\bf Institutional Review Board (IRB) Approvals or Equivalent for Research with Human Subjects}
    \item[] Question: Does the paper describe potential risks incurred by study participants, whether such risks were disclosed to the subjects, and whether Institutional Review Board (IRB) approvals (or an equivalent approval/review based on the requirements of your country or institution) were obtained?
    \item[] Answer: \answerNA{}
    \item[] Justification: \answerNA{}

\end{enumerate}

\end{document}

%% file: sections/introduction.tex
 \begin{wrapfigure}{R}{0.5\textwidth}
    \includegraphics[width=\linewidth]{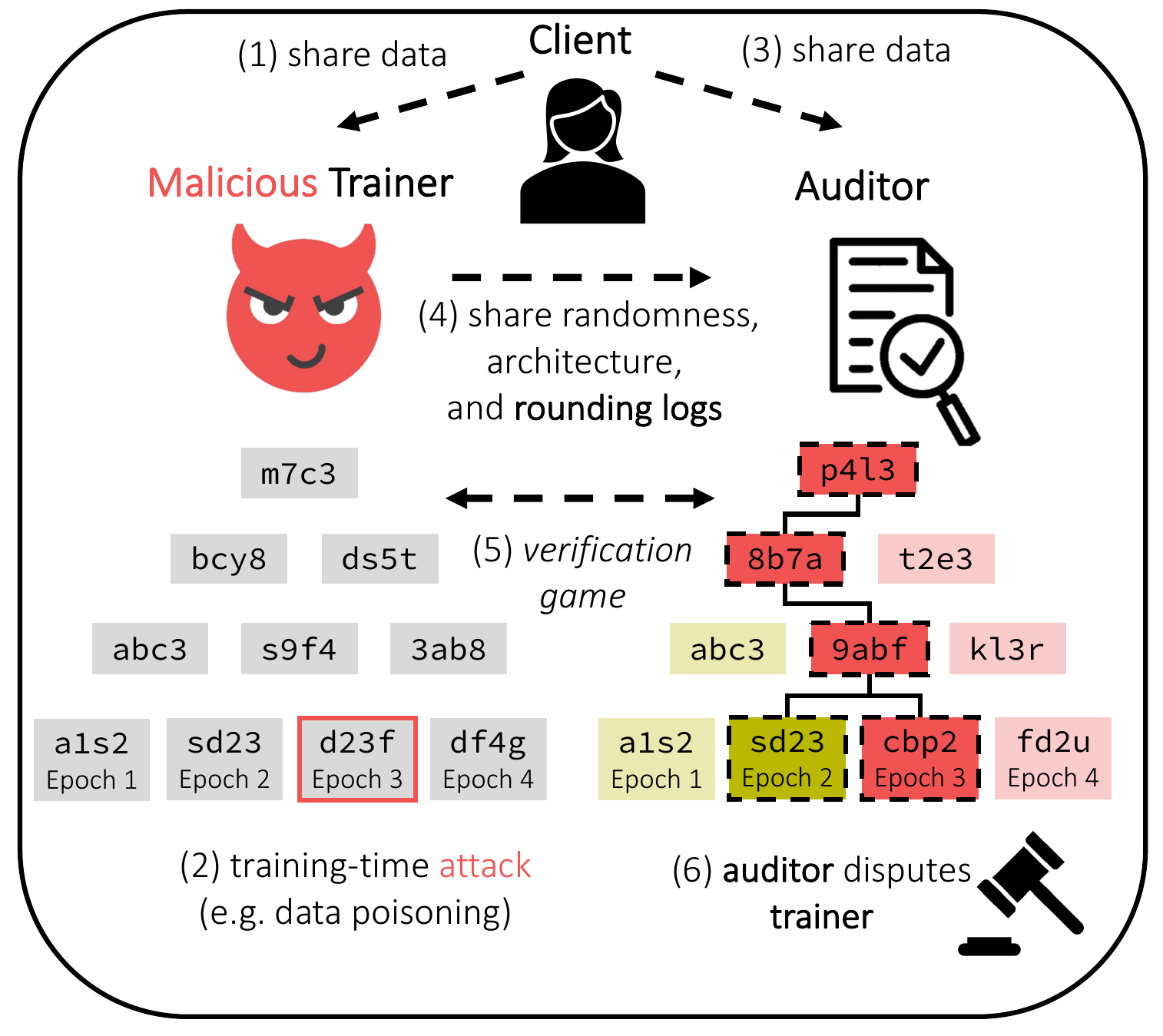}
    \vspace{-0.4cm}
    \caption{Overview of our scheme. After an $\mathrm{auditor}$ challenges a $\mathrm{trainer}$, they train the model, storing weights in a Merkle tree, and enter a binary search procedure to identify the exact steps of the dispute. We show how to control GPU nondeterminism between $\mathrm{auditor}$ and $\mathrm{trainer}$, expanding the set of potential auditors.}
    \label{fig:overview}
\end{wrapfigure}
We are currently in the ``large-scale era'' of machine learning (ML), where the exciting capabilities of modern AI systems have required a dramatic increase in training compute needs  \citep{sevilla2022compute}.
In turn, several model training services, such as Replicate, OpenAI's Finetuning API,  Together AI, Amazon Sagemaker, MosaicML Training, and Gensyn, have been created to support clients who lack the resources to train a model themselves. However, these services require clients to place a significant degree of trust in them to train the model correctly, without introducing a training-time attack such as data poisoning or undetectable backdoors \citep{wan2023poisoning, goldwasser2022planting}. How can we help a client, such as an individual or a small company, hold the service provider accountable in case of misbehavior during training? 

Consider an education start-up that wishes to finetune the Llama-70b language model (70B parameters) on their own curated dataset to support student learning. This task requires significant resources, and the company might even lack the necessary expertise. Instead, they might choose to pay a $\mathrm{trainer}$ with vast computing resources to perform the training  task (Figure \ref{fig:overview}). However, what if the $\mathrm{trainer}$ adds data points that spread misinformation, introduces a backdoor that advances a political agenda for specific prompts, or  tries to save work by under-training the model? If the client starts to notice suspicious model behavior, is there any action they can take? 
We study this problem of \textit{verifiable training}, or ensuring that the training of an ML model was performed correctly.

One possibility is for the $\mathrm{trainer}$ to provide the client with a cryptographic proof that the model was trained according to the specification. However, proof-based systems require cryptographic techniques that can be difficult to scale to the complexity of real-world ML systems. For instance, recent work based on zero-knowledge proof systems for verifiable \textit{inference}, a much simpler task than training, requires more than 8 minutes to generate proofs for only 20 images \citep{liu2021zkcnn}. Thus, practical proof-based methods for verifiable training have only been implemented for simple tasks such as logistic and linear regression \citep{garg2023zk, ames2022ligero}.

An alternative ``optimistic'' approach is to consider a third-pary $\mathrm{auditor}$ (Figure \ref{fig:overview}). This could be a trusted 3rd party, such as a non-profit organization that may not have sufficient computing resources to provide training as a service beyond auditing, or a different provider that the client approaches and wishes to compare with the original model trainer. When a $\mathrm{client}$ suspects foul play, they can ask the $\mathrm{auditor}$ to challenge the $\mathrm{trainer}$ by training the model using the $\mathrm{auditor}$'s own compute, and demonstrate that the $\mathrm{trainer}$ did not train correctly. Based on the provided evidence required from the $\mathrm{auditor}$ (i.e. the precise timesteps model training diverged, as shown in Figure 1), the $\mathrm{client}$ can then choose to refuse the $\mathrm{trainer}$’s model, pursue legal action against the $\mathrm{trainer}$, or even dispute a potentially corrupt $\mathrm{auditor}$ if the client deems such evidence as incorrect, or another $\mathrm{auditor}$ disagrees. This protocol can be efficiently carried out using techniques from the literature on verifiable computing, such as the ``verification game'' method of \citet{teutsch2019truebit}, which uses an interactive binary-search procedure to identify the exact intermediate computation step (e.g. training epoch) where the two parties diverged. Applying verifiable computation techniques to model training is particularly important given the increase in decentralized machine learning services like Gensyn, which seek to make ML compute more accessible by creating a network of many untrusted GPUs.

Unfortunately, the issue with such ``optimistic'' approaches is nondeterminism during model training: two models trained on different GPU types, even with same data order and random seed, can  learn different weights (Figure~\ref{fig:nondeterminism}). The $\mathrm{auditor}$ cannot simply compare their model weights with the $\mathrm{trainer}$'s, and recent work has shown that protocols based on comparing model weights, such as \citet{jia2021learning}'s ``proof of learning,'' are not robust and can be forged due to errors from nondeterminism \citep{thudi2022unlearning, fang2023proofoflearning}.   

Our work addresses this limitation by asking: can the $\mathrm{trainer}$ provide additional information to the $\mathrm{auditor}$ that removes the effects of hardware nondeterminism? Our starting point is the observation that hardware nondeterminism occurs due to the accumulation of errors from floating point operations. For example, a matrix-vector multiply often results in different floating point values on different GPUs, since GPUs often accumulate in different orders.  To address this issue, a natural approach is to perform training using a \textit{higher} precision (e.g. FP32) than the target precision of the model weights (e.g. FP16), and periodically round back to the target precision. The hope is that all floating point errors will be confined to  the higher precision bits, so that the rounded values \textit{are} deterministic. However, this fails because computed values can occasionally straddle the ``rounding boundary'': i.e., the $\mathrm{trainer}$ can round up while the $\mathrm{auditor}$ rounds down, quickly causing them to diverge. Instead, we propose a solution where the $\mathrm{trainer}$  \textit{records} the rounding direction for certain intermediate computation so that $\mathrm{auditor}$ can stay in sync with the $\mathrm{trainer}$.  As this requires the $\mathrm{trainer}$ to record a large number of bits,  we also show how to reduce the amount of data needed to eliminate errors.

We use this strategy to adapt the verification game described by \citet{teutsch2019truebit} for verifiable training. The game's  efficiency lies in our ability to store hashes of model checkpoints in a Merkle tree~\citep{merkle1988merkle}. 
To determine if training was performed according to the specification, the $\mathrm{auditor}$ needs to reconstruct the Merkle tree and compare the resulting Merkle root hash with the Merkle root hash provided by the $\mathrm{trainer}$'s -- if they do not match, the two parties can enter an interactive binary search procedure to identify the exact training step of the dispute. The purpose of the binary search game is to hold both parties accountable: an $\mathrm{auditor}$ should not be able to simply claim that a model was improperly trained, but convince a third-party (e.g., the public, or a $\mathrm{judge}$) by showing at what point during training the $\mathrm{trainer}$ misbehaved. We show our verifiable training scheme can scale to tasks such as full training of ResNet-50 (23M parameters) and finetuning of GPT-2 (117M parameters), significantly outperforming existing methods with respect to both time and storage cost, while eliminating statistical error due to non-determinism. For example, the proposal in prior work \citet{jia2021learning} would require $\mathbf{>140\times}$ more storage cost than our method by comparing model weights at every step in order to achieve low (yet still non-zero) statistical error.

Concretely, our contributions include: 
(1) A method for two parties, training the same model on different GPU types, to achieve identical training results by logging and sharing rounding decisions; 
(2) A verifiable training scheme based on the verification game from \citet{teutsch2019truebit}, which stores model weights in a Merkle tree for efficient comparison between a $\mathrm{trainer}$ and $\mathrm{auditor}$; 
(3) Experiments showing the ability of our approach to scale to large  models such as ResNet-50 and GPT-2 between three different NVIDIA GPU architectures (A40, Titan XP, RTX 2080 Ti); 
(4) Methods to reduce the storage cost of our approach via efficient encoding of rounding logs and an adaptive threshold mechanism to reduce the amount of rounding decisions logged; and 
(5) Comparisons with existing methods, including proof-based systems, that highlight the improved storage and time efficiency of our method.
\footnote{Our method is implemented entirely within the \texttt{pytorch} framework (compatible with version 2.3.1), and is available at \url{https://github.com/meghabyte/verifiable-training}.}

%% file: sections/related_work.tex
Without any verifiable training scheme in place, significant trust is placed in the $\mathrm{trainer}$, leaving a client vulnerable to many different attacks, such as the ``poisoning'' of data samples to cause undesirable behavior (e.g., generating unsafe text \citep{carlini2023poisoning, koh2021stronger, wan2023poisoning}) and planting backdoors triggered by certain inputs \citep{ goldwasser2022planting}. Therefore, training  ML models in trusted environments has been an exciting direction explored by many researchers.  One line of work consists of proof-based systems, where a proof of correctness (for a desired specification) is provided using cryptographic techniques such as succinct non-interactive arguments (SNARKs) \citep{micali1994proofs, SNARKs, lee2020vcnn, liu2021zkcnn, garg2023zk, kang2022scaling}. However, even the most recent proof-based systems for verifiable training suffer extreme latencies, such as 22 minutes for training VGG-11 on one batch of 16 data inputs \citep{abbaszadeh2024dnn}, and have therefore primarily been developed for simpler models (e.g., logistic regression) that are less likely to be delegated to others in the first place
\citep{garg2023zk, ames2022ligero}. Meanwhile,  an alternative solution of training models in a trusted execution environment (TEE), such as NVIDIA's H100, incurs a performance penalty due to the cost of running inside a TEE
\citep{dhanushkodi2024h100}. Furthermore, clients lose all security guarantees if an attacker can extract the attestation key from even one GPU ~\citep{nilsson2020survey, bulck2018foreshadow}. 

Our approach is most similar to proof-of-learning protocols, which consider a trusted 3rd party that compares checkpointing during the course of training  with the original training sequence \citep{jia2021learning}. However, such methods not only incur high storage cost by requiring model weights to be stored frequently, but are non-robust due to errors from training nondeterminism.
Several works have shown that proof-of-learning protocols can be spoofed and fail to verify correctness in several important contexts \citep{fang2023proofoflearning, kong2023membership, thudi2022unlearning}.  Although \citet{choi2023verifying} recently proposed a verification procedure that is immune to several known proof-of-learning attacks, their method is not only limited to supervised learning algorithms, but also based on an assumption that models temporarily overfit data during training, which may not always hold true.

\textbf{GPU Nondeterminism:} 
Prior work has investigated software patches for deterministic training, for instance by enforcing FP accumulation ordering, at a significant cost to efficiency \cite{jooybar2013jooybar, defour2015reproducible, chou2020deterministic, tensorflow2021, zhuang2021randomness}. While these options address deterministic computation on a \textit{single} GPU architecture, achieving deterministic results across multiple GPU architectures remains challenging \cite{crane-2018-questionable, nvidia2022}. We control hardware nondeterminism across GPUs in order to design an efficient and reliable verifiable training scheme. However, our method's impact extends beyond verifiable training, as training nondeterminism can have several negative consequences including bias, reproducibility, and downstream effects on ML pipelines \citep{zhuang2021randomness, crane2018questionable, srivastava2020empirical}.

%% file: sections/setup.tex
\label{sec:setup}

\begin{figure*}
\centering
    \includegraphics[width=\textwidth]{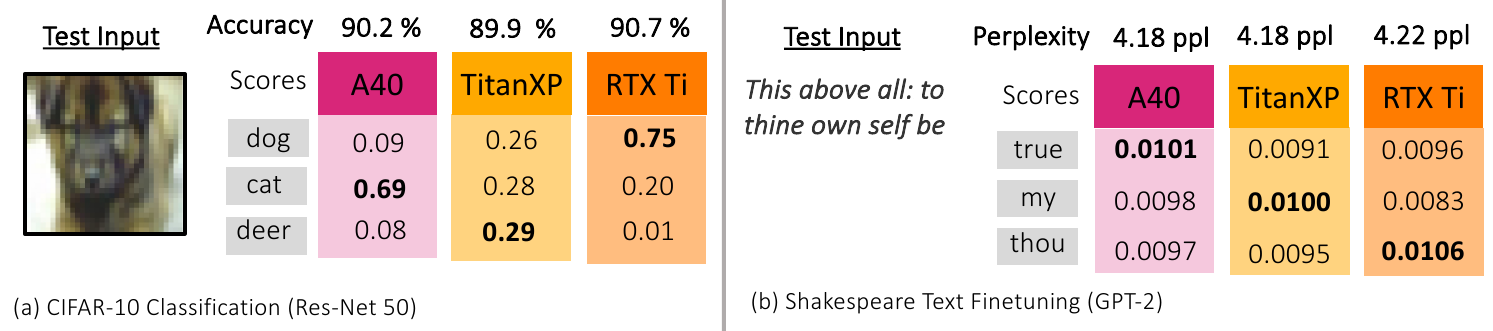}
    \caption{Even after ensuring the same software version, random seed, and use of deterministic algorithms via library flags, training nondeterminism persists between three GPU types.}
    \vspace{-0.2cm}
    \label{fig:nondeterminism}
\end{figure*}

Our method for verifiable training is based on the interactive verification game proposed by \citet{teutsch2019truebit} in the context of blockchains. The core idea is to resolve a dispute between a challenger, in our case the $\mathrm{auditor}$, and a solver, in our case the $\mathrm{trainer}$, for an expensive computation (e.g., model training). In order for the $\mathrm{auditor}$ to take any meaningful action (e.g., pursue legal action), they need to prove the exact source of the dispute (e.g., training time-step where an attack occurred).  If we can save model weights at different time steps into a compact data structure such as a Merkle tree, then identifying the source of disagreement can be done efficiently using binary search \citep{merkle1988merkle}.
More precisely, the verification game consists of the following parties: 
\begin{enumerate}[leftmargin=*,nosep]
    \item $\mathrm{trainer}$, who has putatively trained a model according to a $\mathrm{client}$'s specifications. In our example, this is a service provider with sufficient compute power to train a model.  
    \item $\mathrm{client}$, who receives a model from the $\mathrm{trainer}$ and approaches an $\mathrm{auditor}$.
    \item $\mathrm{auditor}$, who officially challenges the $\mathrm{trainer}$  on behalf of a client. This is a trusted 3rd-party that has sufficient resources but does not necessarily provide training as a service. The  $\mathrm{client}$ can choose several auditors to audit the $\mathrm{trainer}$'s model.  
    \item  $\mathrm{judge}$: Sometimes a $\mathrm{judge}$ may need to arbitrate a legal claim. The $\mathrm{judge}$ can only perform minimal computations (e.g., one training epoch), but can examine the $\mathrm{auditor}$'s claims and enforce a penalty against either the $\mathrm{trainer}$, for incorrect training, or the $\mathrm{auditor}$, for a false alarm. 
\end{enumerate}

When the $\mathrm{trainer}$ is approached by an $\mathrm{auditor}$, they would need to share training parameters, model architecture, and randomness, as shown in Figure \ref{fig:overview}. The $\mathrm{auditor}$ would then replicate the training process, storing model weights in a Merkle tree at the same checkpointing interval as the trainer (every leaf node in a Merkle tree is a hash of the data and every non-leaf node is a hash of its children).  The main loop of the verification game starts when both parties have the root of their respective Merkle trees. If training was performed correctly, then the $\mathrm{trainer}$'s root should match the $\mathrm{auditor}$'s. Otherwise, a binary search procedure is performed,  where the $\mathrm{auditor}$ iteratively descends the 
 Merkle tree until it identifies two consecutive leaf nodes, $i$ and $i+1$, where the hash at $i$ matches that of the $\mathrm{trainer}$, but the hash at leaf $i+1$ does not. This identifies the point in the computation of the dispute.
 
This interactive verification game requires the cooperation of the $\mathrm{trainer}$.  If the $\mathrm{trainer}$ refuses to share the value at a certain node of their Merkle tree within a given time frame, they can be considered to have failed the audit.  Additionally, the  $\mathrm{trainer}$ and $\mathrm{auditor}$ use a Merkle tree to store model weights, requiring far less storage than prior work, if correct training produces identical weights (and identical hash values).The problem is that training nondeterminism leads to weight divergence, and causes this verification game to always fail.  This why we seek to prevent divergence in training.

%% file: sections/challenge.tex
\section{The Nondeterminism Challenge} \label{sec:challenge}

\begin{table*}[]
\centering
\scriptsize
\begin{tabular}{ccccc}
\toprule
\textbf{Example}  & 
\textbf{Sum Order} & 
\textbf{FP32} &
\textbf{Rounded to FP16} \\ 
\midrule
\multirow{2}{*}{
$a,b,c=0.1, -0.1, 0.2$
} 
& $a+b+c$  
& 00111110010011001100110011001101   
& 0011001001100110 \\
& $a+c+b$  
& 00111110010011001100110011001110   
& 0011001001100110 \\

\midrule
\multirow{2}{*}{
$a,b,c=10.02,13.162813186645508,0.2$
} 
& $a+b+c$  
& 01000001101110110001000000000001  
& 0100110111011001  \\
& $a+c+b$  
& 01000001101110110001000000000000   
& 0100110111011000 \\
\bottomrule                                       
\end{tabular}
\caption{Two examples of floating point accumulation error when rounding arithmetic performed higher precision (e.g. FP32) down to lower precision (e.g. FP16). In the second example, the error in the FP32 result transfers to the rounded FP16 result.}
\label{tab:floating_point_a}
\end{table*}

Although there are user-side controls for forcing deterministic operations within a single GPU architecture
, these controls do not prevent nondeterminism between GPU architectures (e.g., NVIDIA H100 and V100), where trained models can have similar aggregate performance (e.g., accuracy) yet yield very different predictions, as shown in Figure~\ref{fig:nondeterminism}~\cite{crane-2018-questionable, nvidia2022}. 
There are three main sources of nondeterminism between GPU types:

\textbf{1. Floating-Point Arithmetic:} 
Computers represent real values using integer and FP representations, typically the IEEE 754 standard (Figure \ref{fig:floats}). There is a tradeoff between the approximation fidelity and the \# of bits used to represent the real values. The chosen precision controls the representable numerical range (e.g., 32-bit FP values can represent values between $1.17549435e-38$ and $3.40282347e+38$). Because computers round to representable FP values, changing the order in which FP values are accumulated can change the resulting sum \cite{kahan1965further, whitehead2011precision}. Over the course of the many operations during training, this can lead to a large difference in the end result between the $\mathrm{trainer}$ and $\mathrm{auditor}$.

\textbf{2. Parallel Computation:} In a GPU, a single operation (called a \textit{kernel}) is executed by thousands of threads in parallel. GPUs contain a set of \textit{streaming multiprocessors} (SMs), which run the \textit{thread blocks} required for the kernel. At the hardware level, these blocks are divided into 
\textit{warps} that are assigned to the available cores. Because different GPUs have a different number and size of compute units,  applications partition arithmetic workloads (e.g., batch matrix multiplies) differently to achieve high performance \cite{nvidia2022}, thus changing the order of FP operations. 

\textbf{3. Memory Hierarchy and Variable Delays:} The time taken for memory access by each thread depends on the physical location of the data, which can create variable delays \cite{jooybar2013jooybar, defour2015reproducible, chou2020deterministic}. The GPU memory hierarchy consists of large amounts of high bandwidth memory (HBM) and small amounts of fast SRAM memory, and maintains an L1 and L2 cache to improve access times. The caches sizes and access times differ across GPU architectures (e.g. an NVIDIA A100 has 192KB / 40 MB of L1/L2 cache memory, while the H100 has 256KB / 50MB). This affects warp scheduling, leading to changes in  operation ordering resulting in nondeterminism. Finally, to compute primitives such as GEMMs ($D = A \cdot B + C$), the workhorse of machine learning, GPUs split the work of computing the tiles of $D$ across a thread block \cite{nvidia2023}, resulting in nondeterminism that a robust verification method needs to control. 

%% file: sections/method.tex
\label{sec:approach}
\begin{figure*}
    \centering
    \includegraphics[width=0.9\textwidth]{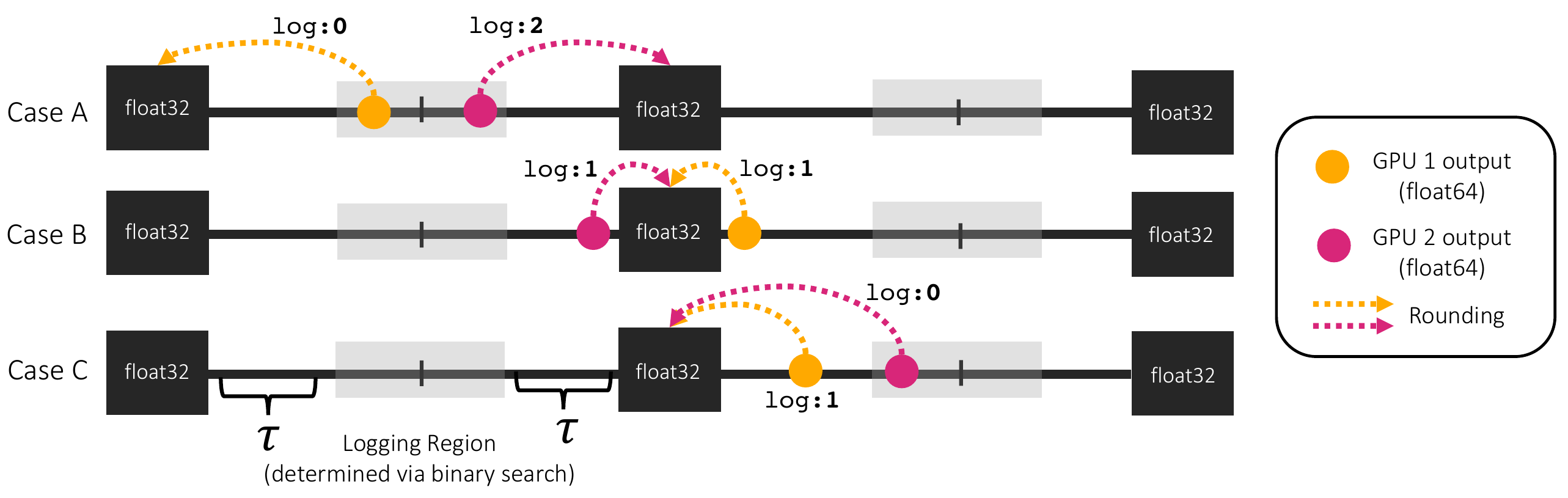}
    \vspace{-0.2cm}
    \caption{Divergence between outputs on two different GPUs (in FP64) for a given function and input can result in different rounding choices when rounding to the nearest FP32. We only wish to log rounding decisions for Case A, where the $\mathrm{auditor}$ should copy the $\mathrm{trainer}$'s rounding choice in order to reach the same value. This requires defining a logging region, determined by a threshold $\tau$,   }
    \label{fig:rounding}
    \vspace{-0.5cm}
\end{figure*}


\subsection{Accumulation Errors Start at Higher Precision Bits}
Our key idea is that if nondeterminism of training between GPU types occurs due to FP operations, then any error will initially be introduced in the lower bits. 
Suppose that both $\mathrm{trainer}$ and $\mathrm{auditor}$ train at  a \textit{higher} FP (e.g., $b_{tr}=64$) precision than the client's target model precision  (e.g., $b_{m}=32$) and then periodically \textit{round} (e.g., $b_{r}=32$) after intermediate computation steps (e.g., a convolution layer). One might hope that this will ``erase" the errors due to nondeterminism, and prevent them from accumulating. Unfortunately, simply rounding to the nearest FP32 after each computation during training is insufficient for determinism.
The problem is due to rounding errors that straddle the \textit{rounding boundary}. Consider Case A in Figure~\ref{fig:rounding}, which shows a divergence in the output of a computation using FP64 on two different GPUs. Because the outputs of GPU 1 and 2 are on different sides of the boundary, rounding to the nearest FP32  results in different values, introducing error. 

What if the $\mathrm{trainer}$ records their rounding choice (e.g., up, down, none) for every intermediate computation? The $\mathrm{auditor}$ could then  copy the $\mathrm{trainer}$'s choice, and therefore round to the exact same value and successfully control for nondeterminism. However, the  $\mathrm{auditor}$ should not copy the  $\mathrm{trainer}$'s behavior for every output (see Cases B \& C, Figure~\ref{fig:rounding}). If a computation output on GPU~1 is too close to the rounded value, then it is possible that GPU~2 is also close in distance but from the opposite direction. In this case, the $\mathrm{auditor}$ should ignore the $\mathrm{trainer}$'s choice. We therefore need to introduce a threshold $\tau$ under which the $\mathrm{trainer}$ does not record their rounding choice.  

Our method requires upper bounding  the divergence  $d_{div}$ between any two different GPUs for any intermediate computation $f$ (i.e. difference in outputs for the same input).
Let $\epsilon_b$ represent the distance between two FP32 values, after rounding to $b_r$ bits of the mantissa (Figure \ref{fig:floats}) and controlling for the exponent. We need to select $b_r$ and $\tau$ such that  $d_{div} < \epsilon_{b_r}$ and $d_{div} < 2\tau$ (Figure \ref{fig:rounding}). Because the set of possible FP numbers is finite, there exist optimal bounds for $b_r$ and $\tau$. In practice, we find that $b_r \leq 32$ and $\tau > 0.25 \cdot \epsilon_{32}$ are sufficient  for standard intermediate computations in neural network training  (e.g., convolution, layer norm) in FP64. We study different values for $b_r$ in Section~\ref{sec:experiments}.

\subsection{Primitives}
We assume both $\mathrm{trainer}$ and $\mathrm{auditor}$ train models using the IEEE-754 standard FP numbers (Figure~\ref{fig:floats}). Besides requiring $\mathsf{read}$ and $\mathsf{write}$ disk I/O operations, we define the following functions: 
\begin{enumerate}[leftmargin=*,nosep]
    \item $\mathsf{rnd}_{b_r}(x)$: rounds input $x$ to the nearest FP up to $b_r$ bits of the mantissa, as shown in Figure \ref{fig:floats}.
    \item $\mathsf{log}(x, b_r, \tau, f)$: logs to file $f$ a logging direction $c$, which is either \texttt{0} (down), \texttt{1} (ignore), or \texttt{2} (up) depending on threshold $\tau$ and rounding amount $b_r$, as shown in Algorithm \ref{alg:Log}.
    \item $\mathsf{rev}(x, b_r, c)$:  reverses rounding of input $x$ based on logging direction $c$. If $x < \mathsf{rnd}_{b_r}(x)$ \& $c=0$, then return $x$ rounded to the nearest float \textit{below} $x$ with $b_r$ precision. If $x > \mathsf{rnd}_{b_r}(x)$ \& $c=2$, then return $x$ rounded to the nearest float \textit{above} $x$ with $b_r$ precision. Otherwise, do not correct. 
    \item $\mathsf{threshold}(l, b_r, b_{tr})$: identifies the optimal threshold to log rounding directions (\texttt{0} or \texttt{2}) instead of \texttt{1}, which the $\mathsf{rev}$ function ignores, based on the binary search procedure in Section \ref{sec:storage}.
    \item $\mathsf{hash}_{\mathsf{sha256}}(\theta)$: creates a SHA-256 hash of provided model weights $\theta$ (in $b_m$ precision).
    \item $\mathsf{tree}(\textit{leaf}_1, \textit{leaf}_2..., \textit{leaf}_n):$ create a Merkle tree where each \textit{leaf} node is the output of $\mathsf{hash}_{\mathsf{sha256}}(\theta)$ for model weights $\theta$ at a given checkpoint, with a checkpointing interval $k$ \citep{merkle1988merkle}.
\end{enumerate}

\subsection{Training and Auditing}
The $\mathrm{trainer}$'s task begins when a client approaches them with dataset $D$, training specifications (epochs $E$, loss function $\mathsf{loss}$, etc.), and a requested model precision $b_m$.The $\mathrm{trainer}$ can then choose a training precision $b_{tr} > b_m$, a rounding amount $b_r \leq b_m$, and a checkpointing interval $k$ to periodically store small $\mathsf{hash}_{\mathsf{sha256}}(\theta)$ of model weights $\theta$ in a Merkle tree, for efficient comparison with an eventual $\mathrm{auditor}$.  Then, as detailed in Algorithm \ref{alg:trainer}, the $\mathrm{trainer}$ can perform training as normal, but after every intermediate computation (e.g., convolution) perform the $\mathsf{rnd}_{b_r}$ operation on each output. Rounding is applied to computations in both the forward and backward passes. Finally, either using  a fixed threshold $\tau$ or a layer-specific optimal $\tau$ from the $\mathsf{threshold}$ function described in Section \ref{sec:storage}, the $\mathrm{trainer}$ applies $\mathsf{log}$, which logs rounding choices \textit{only for the computations an auditor should copy}. The output of the algorithm includes a rounding log file $F$ and the root of the Merkle tree which, along with the shared randomness R and all training parameters, the $\mathrm{trainer}$ can share with any trusted third-pary $\mathrm{auditor}$ who challenges them. 

After a client approaches them, the $\mathrm{auditor}$ initiates the verification game described in Section \ref{sec:setup}. To avoid penalty, the $\mathrm{trainer}$ must cooperate by sharing the rounding amount $b_r$,  randomness R used in training (e.g., a pseudo-random number generator),  the checkpointing interval $k$, and set of rounding logs $F$.  The $\mathrm{auditor}$ then follows the training procedure and corrects their rounding choice (e.g., up or down) to match those logged in $F$ using the $\mathsf{rev}$ operation, as detailed in Algorithm \ref{alg:audit}  (Appendix). By correcting each rounding mismatch during the course of training, the $\mathrm{auditor}$ is able to prevent nondeterminism errors from accumulating. Therefore, the $\mathrm{auditor}$ can store the $\mathsf{hash}_{\mathsf{sha256}}(\theta)$ of model weights $\theta$ in a Merkle tree at interval $k$, knowing that if training was done correctly, the model weights should be identical to the $\mathrm{trainer}$'s at any timestep. The output of Algorithm \ref{alg:audit} is the root of the $\mathrm{auditor}$'s Merkle tree, which they can use to compare with the $\mathrm{trainer}$'s root.

\subsection{Reducing storage cost}
\label{sec:storage}
Logging rounding decisions for every neural network layer output during training incurs a large baseline storage cost, and is our main limitation. For dataset $D$, batch size $B$, training epochs $E$, and model layers $L_\theta$, the upper bound on the total storage cost for verifiable training with our method is: 
\vspace{-0.5em}
\begin{equation}
\text{storage cost (B)} = |D| \times E \times B \times (\sum_{l=1}^Lo_{l,f}+\sum_{l=1}^Lo_{l,b})
\end{equation}
where $o_{l,f}$ and $o_{l,f}$ represent the size of outputs of the forward pass and backward pass of layer $l$. 
Note that the log entries do not need to be kept around in the RAM and can be written straight to the disk. Moreover, this cost is a one-time cost incurred by the $\mathrm{trainer}$, who in our context is likely to be a powerful commercial provider with access to such storage capacity. Furthermore, as we later show in Section \ref{sec:experiments}, for models with many linear layers like Transformer-based language models (e.g., GPT-2), where parameters significantly outnumber intermediate computations, this storage cost is significantly smaller than alternative approaches that require saving model weights \citep{jia2021learning}. Nevertheless, we now describe our method for reducing  storage cost by (i) efficiently encoding rounding logs and (ii) adaptive selection of the threshold $\tau$ to reduce the storage costs.

\textbf{Efficient Encoding:} Each log entry is a value from  the set ${0, 1, 2}$,
as opposed to the FP model weights. We pack sub-sequences of five log entries into a single byte via a fast GPU-based radix-3 to radix-2 conversion, yielding 1.6 bits/entry storage that is close to the best possible packing of 1.58 bits/entry, and yields a 77\% storage reduction relative to naively storing one log entry per byte.

\textbf{Adaptive Threshold:}
Recall that we need to select a threshold $\tau$ that controls for whether the $\mathrm{trainer}$ logs a rounding choice, or instead logs \texttt{1} which the $\mathrm{auditor}$ ignores. The more one can increase $\tau$,  the more \texttt{1} values are recorded, which can make rounding logs more compressible (due to long sequences of  \texttt{1}s). Furthermore, it is possible that the divergence $d_{div}$ between outputs on two different GPUs, given the same input, is function-specific. For example, while convolution requires several matrix multiplications that might result in a large FP accumulation error, normalization operations are unlikely to result in large $d_{div}$, and a larger $\tau$ can be applied. We develop an efficient algorithm (Algorithm~\ref{alg:threshold} in the Appendix) to 
find the optimal value for $\tau$ given a particular layer and data of output values that led to different rounding choices between any two GPUs (e.g., Case A in Figure \ref{fig:rounding}). 
For a given rounding amount $b_r$ and training precision $b_{tr}$, the algorithm performs a binary search between $\tau = 0.25 \cdot \epsilon_{32}$ (our upper bound on the $d_{div}$ between two GPUs for any function) and $\tau = 0.5 \cdot \epsilon_{b_r}$ (the rounding boundary).  By performing this procedure for the different intermediate computations in a model, the $\mathrm{trainer}$ can hope to better compress the rounding log $F$. 

\textbf{Merkle Tree Storage:}
Storing SHA-256 hashes of model weights during training in a Merkle tree creates an efficient mechanism for the verification game described in Section~\ref{sec:setup}, with negligible storage requirements. The audit ends when either the $\mathrm{trainer}$ withdraws, the $\mathrm{auditor}$ confirms that training was performed correctly, or the $\mathrm{auditor}$ can present paths to the two leaves of their Merkle tree where divergence starts, providing evidence to dispute the $\mathrm{trainer}$. 

%% file: sections/experiments.tex
\label{sec:experiments}
\begin{figure*}
    \centering
    \includegraphics[width=0.95\textwidth]{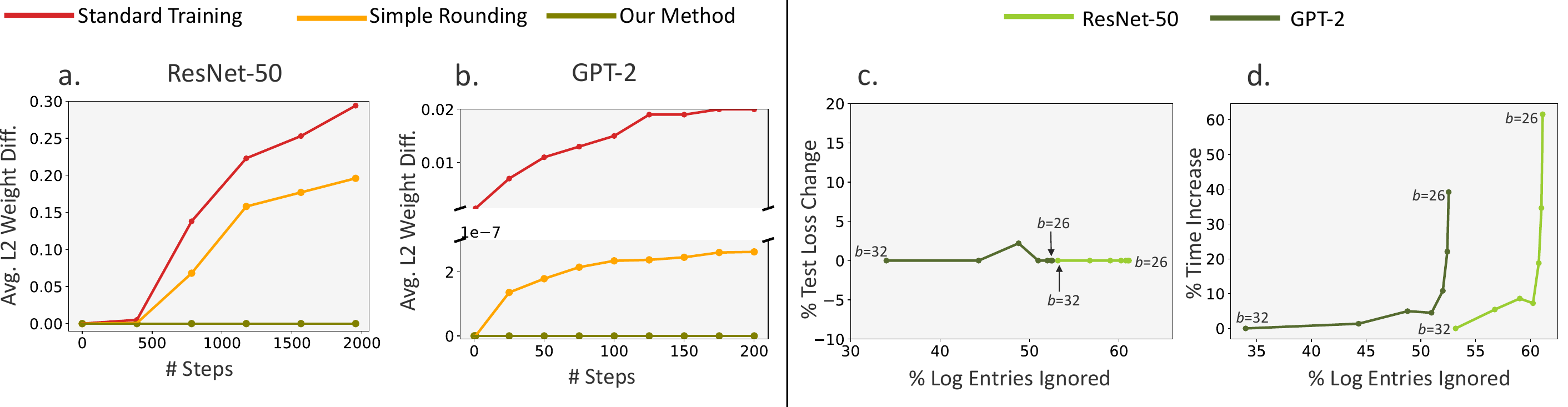}
    \vspace{-0.2cm}
    \caption{ We successfully control for  nondeterminism between GPU types for both ResNet-50 (a.) and GPT-2 (b.) tasks, while standard training and simple rounding without performing $\mathsf{rev}$ corrections result in model divergence over the course of training. Stronger rounding has minimal affect to model performance (c.),  but at the cost of increasing time for $\mathrm{trainer}$ (d.).}
    \label{fig:results}
    \vspace{-0.2cm}
\end{figure*}

\begin{table*}[t]
\footnotesize
    \centering
    \caption{ Efficient encoding reduces storage requirements by 77\%, and rounding to $b=26$ improves the compression further between 5-20\% (values reported for 1 step of training). The original proof-of--learning protocol from \citet{jia2021learning} requires storing 2.78 GB of model weights for GPT-2, or more than \textbf{140x} our storage cost, while still incurring statistical error. }
    \vspace{0.1cm}
    \begin{tabular}{|c|c|c|c|c|c|}
     \hline
         & \textbf{ResNet-50 $b=32$} & \textbf{ResNet-50 $b=26$}  & \textbf{GPT-2 $b=32$} & \textbf{GPT-2 $b=26$} \\  \hline
         Naive Encoding & 456 MB & 456 MB  & 92 MB & 92 MB\\ \hline
         Efficient Encoding & 105 MB & 105 MB & 22 MB & 22 MB\\  \hline
         + Zip Compression & 96 MB &  91 MB & 20 MB & 18 MB\\  \hline
    \end{tabular}
    \label{tab:compression32}
    \vspace{-0.1cm}
\end{table*}

\begin{table*}[t]
\footnotesize
    \centering
    \caption{ Average \# of $\mathsf{rev}$ corrections performed by $\mathrm{auditor}$ per training step. Even at $b=32$, auditing only requires 20-25 corrections (\textbf{2e-6 to 9e-6\%} of samples) per training step.}
    \vspace{0.25em}
    \begin{tabular}{|c|c|c|c|c|c|c|c|}
     \hline
      \textbf{ResNet-50} & \textbf{$b=32$} & \textbf{$b=31$ }& \textbf{$b=30$} & \textbf{$b=29$} & \textbf{$b=28$} & \textbf{$b=27$} & \textbf{$b=26$} \\ \hline
      Forward & $15 \pm  3$ & $6 \pm  2$ & $3 \pm  1$ & $3 \pm  1$ & 0 & 0 & 0 \\ \hline
      Backward & $10 \pm 0.6$ & $6 \pm  0.6$ & $2 \pm  1$ & $0.7 \pm  0.7$ & $0 \pm 0$& $0 \pm 0$ & $0 \pm 0$ \\ \hline \hline
      \textbf{GPT-2} & \textbf{$b=32$} & \textbf{$b=31$} & \textbf{$b=30$} & \textbf{$b=29$} & \textbf{$b=28$} & \textbf{$b=27$} & \textbf{$b=26$} \\  \hline
      Forward & $2 \pm  0.7$ &  $2.3 \pm  0.8$ &  $2.2 \pm  0.4$  &  $0.2 \pm  0.2$ &   $0.4 \pm  0.2$ &   $0 \pm  0$  &   $0 \pm  0$\\ \hline
      Backward & $19 \pm  13$ & $0.75 \pm  0.3$ &  $1.2 \pm  0.4$ &  $0.2 \pm  0.2$ &  $0. \pm  0.0$ &   $0 \pm  0$ &   $0 \pm  0$ \\ \hline
         
    \end{tabular}
    \label{tab:corrections}
    \vspace{-0.5cm}
\end{table*}

\begin{table*}[t]
\footnotesize
    \centering
    \caption{Adaptive thresholds identified for different operations using Algorithm \ref{alg:threshold} with $b=32$. }
    \begin{tabular}{|c|c|c|c|c|}
     \hline
         & \textbf{2D Convolution} & \textbf{Batch Norm} & \textbf{Linear} & \textbf{Layer Norm} \\  \hline
         Dimension & \textit{256 (1,1)} & \textit{(128, 128, 16, 16)} &  \textit{(768,768)}&  \textit{(768,1)} \\ \hline
        $\tau$ & $0.305*2^{-23}$ & $0.499*2^{-23}$ & $0.465*2^{-23}$ & $0.499*2^{-23}$ \\ \hline
    \end{tabular}
    \label{tab:thresholds}
    \vspace{-0.5cm}
\end{table*}

We evaluate our verifiable training method on the two large-scale models listed below with all possible $\mathrm{trainer}$ and $\mathrm{auditor}$ pairs across NVIDIA GPUs  A40, TITAN Xp, and RTX 2080 Ti (see Appendix~\ref{sec:gpus} for more details). In  Section~\ref{sec:comparison}, we compare our method with recent proof-based systems.

\begin{enumerate}[leftmargin=*,nosep]
    \item \textbf{ResNet-50}: We train (from random initialization) ResNet-50 (23M) on CIFAR-10 with dataset size 50K \& batch size B=64. Test accuracy = 90.7\% after 100 epochs training on Titan RTX Ti.
    \item \textbf{GPT-2}: We finetune GPT-2 (117M) on a corpus of Shakespeare text with dataset size 1.1M tokens, batch size B=8, and sequence length 64. Perplexity = 4.22 after 1 epoch training on Titan RTX Ti. 
\end{enumerate}

Figure~\ref{fig:nondeterminism} shows that nondeterminism due to GPU architecture exists for both tasks. While we can repeatedly obtain identical results across training runs on the same GPU architecture, training on different GPU architectures results in fundamentally different models.


\subsection{Implementation and Findings} 
We implement our verifiable training method entirely on top of the \texttt{pytorch} framework,  with \texttt{torch} version 1.13.1 and CUDA version 11.7. The intermediate computations we apply $\mathsf{rnd_b}$ to are layers (e.g., \texttt{torch.nn.Conv2D}) in the model's computation graph. Rounding-related operations ($\mathsf{rnd}$ and $\mathsf{rev}$) either using casting or FP functions (e.g., \texttt{torch.nextafter}) that can run on the GPU, thus having little impact on computational speed. Because we observed that the  \texttt{torch.randn} operation used for dropout in GPT-2 is non-deterministic  for long inputs (even for the same seed, see Appendix~\ref{app:rand}), we  implement our own dropout as our method requires shared randomness $R$. 

\textbf{Successful control for non-determinism:} 
Our method  completely eliminates non-determinism between full training runs of both for both the ResNet-50 training and GPT-2 fine-tuning tasks across all possible $\mathrm{trainer}$ and $\mathrm{auditor}$ pairs between the A40, Titan XP, and RTX 2080 Ti GPUs. As Figure~\ref{fig:results} shows, standard FP32 training results in an increasing divergence (l2-distance of weights) between models on different GPUs over the course of training. Furthermore, we show the simple approach of training in FP64 and rounding to FP32 after every intermediate computation, but without the $\mathrm{auditor}$ correcting rounding decisions with $\mathsf{rev}$, fails to mitigate this issue. Only our method, in which the $\mathrm{auditor}$ follows the rounding decisions ($b_r=32$) made by the $\mathrm{trainer}$ for every intermediate computation, eliminates non-determinism and persists over time. 
Our implementation, which requires disk I/O during training to store the rounding decisions, results in a small increase in training time for the $\mathrm{trainer}$ (1.2-1.4x) and  $\mathrm{auditor}$ (1.3-1.7x) using a non-optimized, protoype implementation (Table \ref{tab:time}). We report the storage requirements of our method in Table \ref{tab:compression32}, showing that our efficient encoding scheme reduces the size of the $\mathrm{trainer}$'s rounding logs by 77\%, relative to naive logging.  
Because the Merkle tree stores 32-byte SHA-256 hashes, its overall size (KBs) and creation time are negligible and not reported. 
Finally, we show that decreasing the rounding amount $b$ to values even as low as 26 has little effect on model performance (we observe no change in accuracy, so report test loss), but increase training time (Figure \ref{fig:results}). We observe that smaller values of $b$ do allow more log entries to be ignored, improving compression of the file, which we discuss next. 

\textbf{Compression with adaptive threshold:}
Our approach  outperforms (Table~\ref{tab:compression32}) the storage costs of proof-of-learning protocols that save model weights for GPT-2 (2.78GB), which has many linear layers -- we observe more than \textbf{140x} reduction relative to the approach in \citet{jia2021learning}. We further reduce the storage cost of our method by decreasing the rounding amount $b$ and implementing the adaptive thresholding strategy (Section~\ref{sec:storage}).
Table~\ref{tab:thresholds} reports adaptive thresholds $\tau$ for four different \texttt{pytorch} layers at rounding amount $b_r=32$. Convolutions require the lowest $\tau$, indicating larger divergence in outputs between GPU types, which is expected due to the large \# of matrix multiplications.  Meanwhile, $\tau$ is higher for normalization layers, likely due to smaller divergences between GPU types. Because adaptive thresholding seeks to reduce the \# of times rounding decisions (\texttt{0} and \texttt{2}) are logged and improve log file compression, we report storage cost after zip compression in Table \ref{tab:compression32}. As expected, more aggressive rounding (which results in a higher $\tau$) improves the compression rate. Although the compression gains are mild in comparison to our encoding step, they build-up over the course of training.
Finally, we report the average \# of  $\mathsf{rev}$ corrections an $\mathrm{auditor}$ needs to perform for one training step in our two tasks (Table \ref{tab:corrections}). These values are surprisingly small in comparison to the \# of operations logged --  only a maximum of  \textbf{2e-6\%} (ResNet-50) and \textbf{9e-6\%}  (GPT-2) of logged values, are actually needed by the $\mathrm{auditor}$!   We also observe that severe rounding (e.g., $b=27$) completely eliminated the hardware non-determinism for our tasks, requiring no corrections from the $\mathrm{auditor}$.
This shows a huge gap between the \# of values currently saved by the $\mathrm{trainer}$ and those needed by the $\mathrm{auditor}$, motivating an exciting future possibility of significantly reducing the storage cost of our method if we could reliably predict when a divergence will not occur.
 
\subsection{Comparison with alternative approaches} \label{sec:comparison}

\textbf{Logistic Regression:}
\citet{garg2023zk} recently proposed a zero-knowledge proof-based system for verifiable training of a logistic regression, which importantly does not leak information about the client's data or require a trusted third-party $\mathrm{auditor}$, unlike our work. However, since verifiable training itself is motivated by a client not having sufficient resources to train the model, it is crucial to consider the implications of scale. The authors report the prover time and proof size requirements for one training pass of logistic regression on a dataset of $2^{18}$ items, with 1024 dimensions and a batch size of 2014, as \textbf{72 seconds} (training and proof generation time) and \textbf{350 MB} respectively. We replicate this training task, and find that our method significantly improves upon both  storage and time requirements, requiring only \textbf{106 KB} and \textbf{7 seconds} (both training and auditing). Furthermore, because \citet{garg2023zk} do not report the duration of  ``offline phase'' of their method, their reported value is a lower bound on the actual time required. Finally, we note that the original proof-of-learning protocol from \citet{jia2021learning}, which also considers a trusted third-party, would require \textbf{9.2 MB per training step} to store all model weights. Therefore, our method is at least \textbf{85x} more space efficient.

\textbf{VGG-11:} Concurrent to this work, \citet{abbaszadeh2024dnn} introduce a zero-knowledge proof-of-training protocol for deep neural networks, presenting results for one batch step of training for a simplified version of the VGG-11 model with 10M parameters, which is less than the original VGG-11 network and ResNet-50 \citep{simonyan2015deep}. While the authors do not provide 
architectural details, we can assume that increasing the \# of parameters to the original VGG-11 would only increase their reported proof time and size.  We, therefore, compare their reported values with an implementation of our method for the same task of verifying the training of VGG-11 on CIFAR-10 with a batch size of 16. While their use of incrementally verifiable computation leads to tractable proof size (1.36MB vs. the 1.2MB per iteration cost of our method), \citet{abbaszadeh2024dnn}'s method requires \textbf{22 min. per training iteration}. In comparison, our method requires training and auditing times of only 6 sec. per iteration and is significantly more efficient (factor of \textbf{220x}), an important consideration for model training as a commercial service.

Finally, in Appendix Section \ref{app:gpt2inference}, we compare our results with an adaption of \citet{gupta2023gpt}'s protocol for secure inference of GPT-2. Compared with our method's storage cost (18MB) and training time (11s for training, 13.5s for auditing), scaling \citet{gupta2023gpt}'s protocol for training would introduce around a \textbf{10,000x} data and \textbf{40x} time overhead. While proof-based systems provide strong security guarantees without a third party, they do so at the cost of relying on hard-to-scale cryptographic techniques, as well as approximating non-linear functions that can harm performance. 

%% file: sections/security.tex
Our work makes a \textit{1-of-n} honesty assumption, i.e., as long as one of $n$ auditors is honest, any attack from a malicious trainer that results in diverging model weights will be detected. One consideration is the potential manipulation of the rounding logs by an adversarial $\mathrm{trainer}$ who could select rounding decisions that achieve a desired outcome, and which the $\mathrm{auditor}$ would follow. Concretely, let us define our threat model so that the $\mathrm{trainer}$ knows an $\mathrm{auditor}$'s GPU a priori.  Recall that an $\mathrm{auditor}$ only copies the $\mathrm{trainer}$'s rounding decision in Case~A in Figure~\ref{fig:rounding}, when both GPUs compute values close to the rounding boundary. Under this threat model, the $\mathrm{trainer}$ can identify the $n$ steps where the $\mathrm{auditor}$ is close to the boundary (as in Case~A), enumerate the set of $2^n$ different models that result from different rounding decisions, and selectively pick a model that exhibits a desired property. 

However, the $\mathrm{trainer}$ cannot use this strategy to embed an arbitrary property (e.g., a specific backdoor). It can only select from the set of models that differ in certain rounding decisions, which all require the $\mathrm{trainer}$ to use the correct training specifications accepted by the $\mathrm{client}$ (such as exact training data \& hyperparameters).  Furthermore, since the expected \# of divergences between the $\mathrm{trainer}$ and the $\mathrm{auditor}$ is extremely small (see Table~\ref{tab:corrections}), the set of possible models where an $\mathrm{auditor}$ would not detect an attack (e.g., many $\textsf{rev}$ ops) is limited.  Finally, we show in Table~\ref{tab:divergence} in the appendix that the divergence (measured both as $\ell_2$-norm between model weights and output distributions) due to GPU non-determinism is significantly less than the divergence due to data ordering during training. Therefore, if a $\mathrm{client}$ will accept a model trained with \textit{any} random ordering of the data during training, then it is unlikely that an adversarial $\mathrm{trainer}$ --- that can only alter rounding decisions --- could produce a model that the $\mathrm{client}$ would not accept. Nevertheless, fully understanding the model properties obtained by manipulating rounding logs adversarially is an important future direction.

%% file: sections/appendix.tex
\section{IEEE Floating Point Image}
\begin{figure}
\centering
\vspace{-0.2em}
    \fbox{\includegraphics[width=0.35\textwidth]{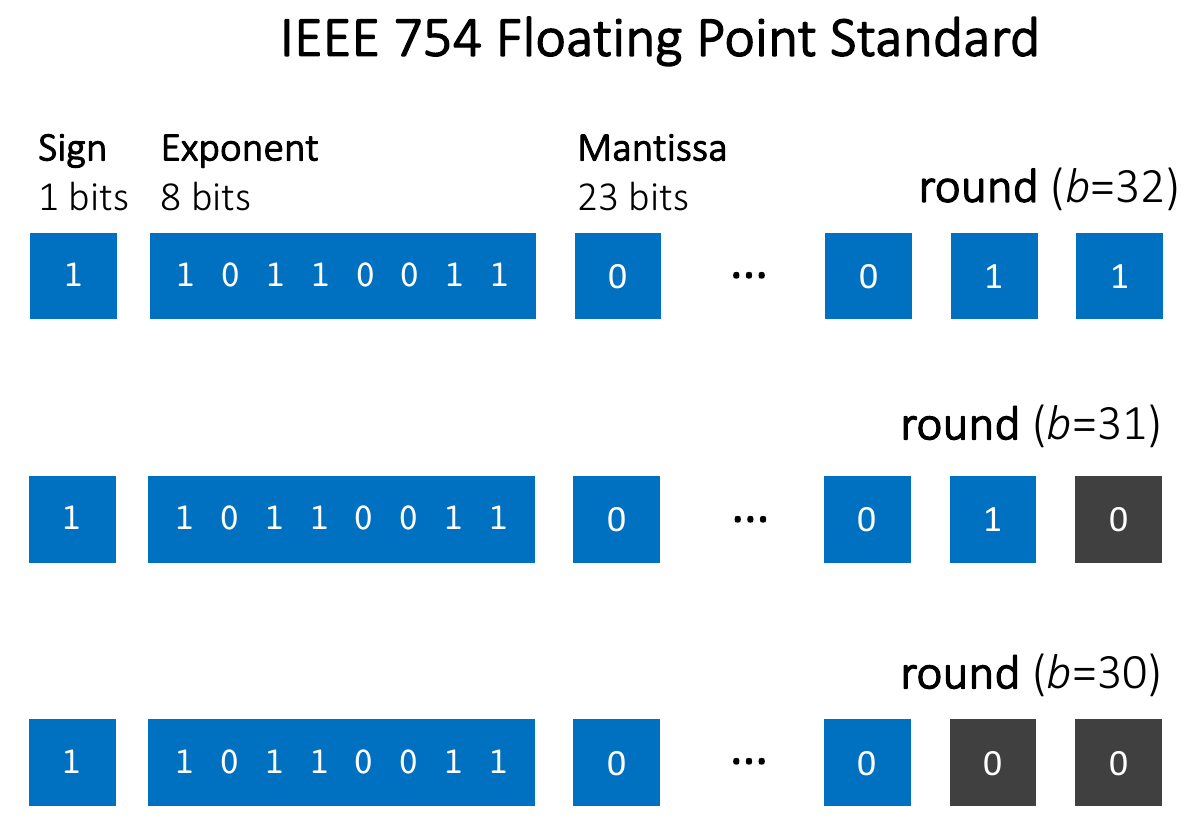}}
    \vspace{-0.2cm}
    \caption{We define rounding to $b$ bits as rounding to the nearest 32-bit FP number that has 0s in the last $32-b$ bits of the mantissa, after accounting for the exponent.}
    \label{fig:floats}
\vspace{-0.5cm}
\end{figure}

See Figure \ref{fig:floats}.

\section{GPU Details} \label{sec:gpus}
All experiments reported in our paper are run with the following three GPUs: 
\begin{itemize}[nosep]
    \item NVIDIA Titan XP: 3840 Cores, 12 GB
    \item NVIDIA RTX 2080 Ti: 4352 Cores, 11 GB
    \item NVIDIA A40: 10752 Cores, 48 GB
\end{itemize}

We are able to successfully replicate training runs between all pairs of these 3 GPUs. 

\section{Logging Algorithm}
See Algorithm \ref{alg:Log}
\section{Train Algorithm}
See Algorithm \ref{alg:trainer}.

\section{Audit Algorithm}
See Algorithm \ref{alg:audit}.

\section{Adaptive Thresholding Algorithm}
See Algorithm \ref{alg:threshold}.

\section{Time Requirements}
See Table \ref{tab:time}.

\section{Model Divergence Comparison}
See Table \ref{tab:divergence}.

\section{Random Number Generation} \label{app:rand}
Our verifiable training scheme requires shared randomness between the $\mathrm{trainer}$ and $\mathrm{auditor}$, which is used for deciding input data batching, weight initialization, and  operations such as dropout (randomly setting outputs to zero). More formally, our scheme requires sharing the same random seed and pseudo-random generator. However, in our implementation based on \texttt{pytorch} (assuming the same software version between $\mathrm{trainer}$ and $\mathrm{auditor}$), we chose to rely on the the \texttt{torch} random seed functionality. While this successfully controls for batch input ordering and weight initialization, it is unfortunately not sufficient for random number generation, as operations such as \texttt{torch.nn.randn()} leverage parallelism when the requested \# of values is higher than a certain amount.  Specifically, we found that across T40, RTX 2080 Ti, V100, A40, and A100, given the same seed, torch.randint() produces identical tensors onlt up to size 40960. At size 40961, T40 (which is an older GPU) deviated from the rest. Likewise, at size 69633, 2080 Ti deviated from the rest, and so on. Based on these observations, we arranged for calls to torch.randint() in the dropout layer (which is the only operation using large random tensors in our tasks) to be replaced by generating and concatenating multiple random tensors of size 40960 or less. Specifically, a random tensor of size $n  > 40960$ is generated by concatenating ($n // 40960$) random tensors of size 40960 and one random tensor of size ($n \% 40960$). However, we emphasize that it is therefore important in our scheme either for both parties to implement this change a priori, or simply use an external source for pseudorandomness.

\section{Comparison with GPT-2 Inference} \label{app:gpt2inference}
The previously discussed proof-based systems for verifiable training by-pass the need for a third-party $\mathrm{auditor}$, but very few efficient systems exist in the literature. Many more works study secure \textit{inference} of deep neural networks, which could be used to construct verifiable training protocols with stronger security guarantees than ours (e.g., allowing a $\mathrm{trainer}$ to keep a proprietary model's weights private), but come at a significant cost to performance and resources. 
To demonstrate this, we consider adapting \citet{gupta2023gpt}'s protocol for secure inference of GPT-2 based on multi-party computation, to our context of verifiable training. \citet{gupta2023gpt} show how  two parties,  the $\mathrm{client}$ with private data and the $\mathrm{trainer}$, can jointly compute the forward pass of a known model architecture without revealing additional information beyond the model output to each other. Because they report the the communication overhead $P=0.37 $GB and time $T = 0.96 $ seconds for one forward pass on a single data input, we can calculate $2 \times P \times D \times E = \textbf{ 189 GB}$ and $2 \times T \times D \times E =$ \textbf{ 983 seconds} as estimated communication cost and time, respectively, for 1 step of training in out GPT-2 task, where 2 considers both the forward and backward pass. Compared with our method's required storage cost (18MB) and training time (11s for training, 13.5 seconds for auditing), scaling \citet{gupta2023gpt}'s protocol for training would introduce around a \textbf{10,000x} data and \textbf{40x} time overhead.

\begin{algorithm}
\caption{$\mathsf{train}$}
\label{alg:trainer}
\vspace{1mm}
\textsc{Input:}  dataset $D$, epochs $E$, batch size $B$, shared randomness $R$, model $W_\theta$, loss function $\mathsf{loss}$, rounding amount $b_r$, training precision $b_{tr}$, target model precision $b_m$, checkpointing interval $k$  \\
\textsc{Output:} Merkle tree root $M_{root}$, rounding log file $F$ \\
\begin{algorithmic}[1]

\STATE $F, M_{leaves} \leftarrow$ create  empty file and leaf list
\STATE $W_\theta \leftarrow \mathsf{init}(R, b_{tr})$ \texttt{//initialize weights}
\STATE $T \leftarrow \frac{D*E}{B}$
\FOR{$ t=1...T$} 
\STATE $input  \leftarrow \mathsf{batch}(R, D, B)$ \texttt{// get data batch}

\texttt{// forward pass}
\FOR{layer $l_{\theta} \in W_\theta.\text{layers}$ }
\STATE  $ output \leftarrow l_{\theta} (input)$
\STATE  $ \tau \leftarrow \mathsf{threshold}(l_{\theta}, b_r, b_{tr})$ \texttt{//set threshold}
\STATE $\mathsf{log}(output, b_r, \tau, F)$
\STATE $ output  \leftarrow \mathsf{rnd}_{b_r}(output) $ 
\STATE $ input  \leftarrow output$
\ENDFOR
\STATE $loss \leftarrow \mathsf{loss}(output)$
\STATE \texttt{// backward pass, reversed layers}
\STATE $grad\_output \leftarrow \nabla_{\mathsf{loss}}$
\FOR{layer $l_{\theta} \in W_\theta.\text{layers}$} 
\STATE  $ grad\_input \leftarrow \nabla_{l_{\theta}} (grad\_output)$ 
\STATE  $ \tau \leftarrow \mathsf{threshold}(\nabla_{l_{\theta}}, b_r, b_{tr})$
\STATE $\mathsf{log}(grad\_input, b_r, \tau, F)$ 
\STATE $ grad\_input  \leftarrow \mathsf{rnd}_{b_r}(grad\_input) $ 
\STATE $ grad\_output  \leftarrow grad\_input$
\ENDFOR
\STATE $\theta \leftarrow$ update \texttt{update weights}
\IF{$t$ mod $k = 0$}
\STATE $M_{leaves}.\mathsf{append}(\mathsf{hash}_{\mathsf{sha256}}(\theta\text{ in precision }b_m))$
\ENDIF
\ENDFOR
\STATE $M_{root} \leftarrow \mathsf{tree}(M_{leaves}) \texttt{ // create Merkle tree}$ 
\STATE \textbf{return} $F, M_{root}, $ and model $W_\theta$ in target precision $b_m$
\end{algorithmic}
\end{algorithm}

\begin{algorithm}
\caption{$\mathsf{audit}$}
\label{alg:audit}
\vspace{1mm}
\textsc{Input:}  dataset $D$, epochs $E$, batch size $B$, shared randomness $R$, model $W_\theta$, loss function $\mathsf{loss}$, rounding amount $b_r$, training precision $b_{tr}$, target model precision $b_m$, checkpointing interval $k$, log file $F$ from $\mathrm{trainer}$\\
\textsc{Output:} Merkle tree root $M_{root}$ \\
\begin{algorithmic}[1]

\STATE $M_{leaves} \leftarrow$ create  empty leaf list
\STATE $W_\theta \leftarrow \mathsf{init}(R, b_{tr})$ \texttt{//initialize weights}
\STATE $T \leftarrow \frac{D*E}{B}$
\FOR{$ t=1...T$}
\STATE $input  \leftarrow \mathsf{batch}(R, D, B)$ \texttt{// get data batch}

\texttt{// forward pass}
\FOR{layer $l_{\theta} \in W_\theta.\text{layers}$ }
\STATE  $ output \leftarrow l_{\theta} (input)$ 
\FOR{$output_i \in output$}
\STATE \texttt{// Match trainer rounding}
\STATE $c \leftarrow \mathsf{read}(output_i, F)$ 
\STATE $output_i \leftarrow \mathsf{rev}(output_i, b_r, c)$
\ENDFOR
\STATE $ input  \leftarrow output$
\ENDFOR
\STATE $loss \leftarrow \mathsf{loss}(output)$
\STATE \texttt{// backward pass}
\STATE $grad\_output \leftarrow \nabla_{\mathsf{loss}}$
\FOR{layer $l_{\theta} \in W_\theta.\text{layers}$ }
\STATE  $ grad\_input \leftarrow \nabla_{l_{\theta}} (grad\_output)$ 
\FOR{$grad\_input_i \in grad\_input$} 
\STATE \texttt{// Match trainer rounding}
\STATE $c \leftarrow \mathsf{read}(grad\_input_i, F)$ 
\STATE $grad\_input_i \leftarrow \mathsf{rev}(grad\_input_i, b_r, c)$
\ENDFOR
\STATE $ grad\_output  \leftarrow grad\_input$
\ENDFOR
\STATE $\theta \leftarrow$ update \texttt{update weights}
\IF{$t$ mod $k = 0$}
\STATE $M_{leaves}.\mathsf{append}(\mathsf{hash}_{\mathsf{sha256}}(\theta\text{ in precision }b_m))$
\ENDIF
\ENDFOR
\STATE $M_{root} \leftarrow \mathsf{tree}(M_{leaves}) \texttt{ // create Merkle tree}$
\STATE \textbf{return} $M_{root}$
\end{algorithmic}
\end{algorithm}

\begin{algorithm}
\caption{$\mathsf{threshold}$}
\label{alg:threshold}
\vspace{1mm}
\textsc{Input:}  layer $l$, rounding amount $b_r$, training precision $b_{tr}$  \\
\textsc{Output:}  threshold $\tau$
\begin{algorithmic}[1]
\STATE $P \leftarrow$ initialize empty list
\STATE $N, T \leftarrow$ initialize large \# of data points and iterations
\FOR{i=1...N}
\STATE $GPU1, GPU2 \leftarrow$ select two different GPU architectures
\STATE $x \leftarrow $ select random input for layer $l$ in $b_{tr}$ floating-point precision
\STATE $y_1 \leftarrow l_{GPU1}(x),  y_2 \leftarrow l_{GPU2}(x), $ apply layer $l$ on input $x$ on each GPU
\IF{$\mathsf{rnd}_{b_r}(y_1) \neq \mathsf{rnd}_{b_r}(y_2)$}
\IF{$y_1 > \mathsf{rnd}_{b_r}(y_1)$ and $y_2 < \mathsf{rnd}_{b_r}(y_2)$}
\STATE $P.\mathsf{append}(|y_1-\mathsf{rnd}_{b_r}(y_1)|)$
\STATE $P.\mathsf{append}(|y_2-\mathsf{rnd}_{b_r}(y_2)|)$
\ENDIF
\IF{$y_1 < \mathsf{rnd}_{b_r}(y_1)$ and $y_2 > \mathsf{rnd}_{b_r}(y_2)$}
\STATE $P.\mathsf{append}(|y_1-\mathsf{rnd}_{b_r}(y_1)|)$
\STATE $P.\mathsf{append}(|y_2-\mathsf{rnd}_{b_r}(y_2)|)$
\ENDIF
\ENDIF
\ENDFOR
\STATE \texttt{ //binary search to select threshold}
\STATE $lower, upper, \tau \leftarrow 0.25*(2^{-23}), 0.5*(2^{9-b_r}), 0$
\FOR{t=1...T}
\STATE $\tau \leftarrow (lower + upper)/2$ 
\STATE $success \leftarrow True$
\FOR{$p_i \in P$}
\STATE $\exp \leftarrow $get exponent of $p_i$
\IF{$p_i < \exp*\tau$} 
\STATE $success \leftarrow False$
\ENDIF
\ENDFOR
\IF{$success$}
\STATE $lower \leftarrow \tau$
\ELSE
\STATE $upper \leftarrow \tau$
\ENDIF
\ENDFOR
\STATE \textbf{return} $\tau$
\end{algorithmic}
\end{algorithm}

\pagebreak
\begin{table*}[t]
    \centering
    \caption{Training time requirements, including Merkle tree operations (at $k=5$), for 1 step of training broken down by stage of our verifiable training process. Note that reported times are specific to the particular dataset, batch size, and task, and using a non-optimized prototype codebase -- therefore the relative increase is time is more important.} 
    \begin{tabular}{|c|c|c|}
     \hline
        & ResNet-50 &  GPT-2 \\ \hline
         Original (No Rounding or Disk I/O) &  24s& 8s \\ \hline
         Trainer   & 28s  &  11s \\  \hline
         Auditor  & 31s & 13.5 \\  \hline
    \end{tabular}
    \label{tab:time}
\end{table*}

\begin{table*}
    \centering
    \caption{Comparison of model divergence due to data ordering versus GPU non-determinism. Reported numbers are averaged between 10 pairs of models, error bars are standard deviation.} 
    \begin{tabular}{|c|c|c|}
     \hline
        Metric & Data Ordering & GPU Non-determinism\\ \hline
        l2 weight difference & $133.2 \pm 9$ &   $1.1 \pm 0.07$ \\ \hline
         l2 output distance & $5.3 \pm 0.03$ &   $0.26 \pm 0.02$  \\ \hline
    \end{tabular}
    \label{tab:divergence}
\end{table*}

\begin{algorithm}
\caption{$\mathsf{log}$}
\label{alg:Log}
\vspace{1mm}
\textsc{Input:}  value $x$, rounding amount $b_r$, threshold $\tau$, file $F$  \\
\begin{algorithmic}[1]
\STATE $\exp \leftarrow $ get exponent of $x$ 
\IF{$|x - \mathsf{rnd}_{b_r}(x)| > \exp * \tau$ \AND $x < \mathsf{rnd}_{b_r}(x)$}
\STATE  $\mathsf{write}(2, F)$ \texttt{//  log rounding up}
\ELSIF{$|x - \mathsf{rnd}_{b_r}(x)| > \exp* \tau$ \AND $x > \mathsf{rnd}_{b_r}(x)$}
\STATE  $\mathsf{write}(0, F)$ \texttt{// log rounding down}
\ELSE
\STATE  $\mathsf{write}(1, F)$ \texttt{// log rounding ignore}
\ENDIF
\end{algorithmic}
\vspace{-.25em}
\end{algorithm}